\documentclass[aps,floats]{revtex4} 

\usepackage[dvips]{color}     

\usepackage{epsfig,comment,pifont,rotating,colortbl,multirow,graphicx,mathrsfs,makeidx,color,
longtable,fancyhdr,floatflt,fancybox,calc,amsmath,amsfonts,amssymb,amsthm,latexsym}

\usepackage[english]{babel}
\usepackage[latin1]{inputenc} 

\newcommand{\nc}{\newcommand}
\nc{\be}{\begin{equation}} \nc{\ee}{\end{equation}}
\nc{\bea}{\begin{eqnarray}} \nc{\eea}{\end{eqnarray}}
\nc{\beas}{\begin{eqnarray*}} \nc{\eeas}{\end{eqnarray*}}
\nc{\bt}{\begin{tabular}} \nc{\et}{\end{tabular}}
\nc{\ba}{\begin{array}} \nc{\ea}{\end{array}}
\nc{\ov}{\overline} \nc{\wt}{\widetilde}
\nc{\bvec}{\mathbf} \nc{\dis}{\displaystyle}
\nc{\vecna}{\mbox{\boldmath $\nabla$}}
\nc{\ellbf}{\mbox{\boldmath $\ell$}}
\nc{\pr}{{\rm I}} \nc{\se}{{\rm II}}
\nc{\noi}{\noindent} \nc{\rd}{{\rm d}} \nc{\erm}{{\rm e}}
\nc{\lin}{\noindent\mbox{}\hrulefill \\}
\nc{\gr}{^{\rm o}}

\begin{document}


\title{Delayed luminescence induced by complex domains 

in water and in aqueous solutions} 

\author{C. Colleoni$^{\rm a}$}
\author{S. Esposito$^{\rm b}$}
\author{R. Grasso$^{\rm c}$}
\author{M. Gulino$^{\rm c,d}$}
\author{F. Musumeci$^{\rm c,e}$}
\author{D. Romeli$^{\rm a}$}
\author{G. Rosace$^{\rm a}$}
\author{G. Salesi$^{\rm a,f,}$\footnote{\textit{Corresponding author} \\ e-mail: salesi@unibg.it \\ tel/fax: +390352052077}}
\author{A. Scordino$^{\rm c,e}$} 

\affiliation{\ \\ \mbox{$^{\rm a}$Dipartimento di Ingegneria e Scienze Applicate, Universit\`a di Bergamo, viale Marconi 5, Dalmine, Italy}}


\affiliation{$^{\rm b}$Istituto Nazionale di Fisica Nucleare, Sezione di Napoli, via Cinthia, Naples, Italy}

\affiliation{$^{\rm c}$Laboratori Nazionali del Sud, Istituto Nazionale di Fisica Nucleare, Catania, Italy}

\affiliation{\mbox{$^{\rm d}$Facolt\`a di Ingegneria e Architettura, Universit\`a di Enna ``Kore'', Enna, Italy}}

\affiliation{$^{\rm e}$Dipartimento di Fisica e Astronomia, Universit\`a di Catania, Catania, Italy,}

\affiliation{\mbox{$^{\rm f}$Istituto Nazionale di Fisica Nucleare, Sezione di Milano, via Celoria 16, Milan, Italy}}

\begin{abstract}
\noi 
Many recent studies on water have conjectured a complex structure composed of hydrogen bonded low- and high-density domains. In this work the structure of pure water and aqueous solutions of silica gel (TEOS) has been investigated by using delayed luminescence, which previously have showed significant increase in aqueous salt solutions where low-density domain formation is expected. Photon emission shows an Arrhenius trend with an activation energy in water-TEOS solutions larger than in pure water and salt-water solutions. Moreover, delayed photon emission decay shows an intrinsic lifetime of about $5\,\mu$s both in solutions and in pure water that, along with secondary lifetimes induced by the presence of TEOS, could be related to the formation of different domains.

\end{abstract}

\maketitle


\section{Introduction}

\noi Water is the most abundant substance on Earth and presents a number of anomalous properties which have been related also to the water's role in the chemistry of life. In particular intracellular water has been recognized to behave differently from bulky water, as protein-water interactions take place at protein surface where properties of water change [Wiggins 1990]. The results of both experimental and theoretical works have indicated as water is not simply a solvent but it actively engages and interacts with biomolecules in complex and essential ways.

The strangeness of water is a consequence of the extensive three-dimensional hydrogen bonding of water molecules to one another, which is what makes water a liquid rather than a gas at ordinary temperatures and pressure. In fact, while the molecular movements in liquid water require constant breaking and reorganization of individual hydrogen bonds on a pico-second time scale, at any instant, the degree of hydrogen bonding is very high, more than 95\%.  These strange properties of water have intrigued generations of scientists and many models [Stanley et al. 2008] have been proposed to explain peculiar behaviour and anomalies of water. It is evident that liquid water is not a homogeneous structure at the molecular level, showing a dynamic equilibrium among changing percentages of assemblages of different oligomers and polymer species (clusters), whose structure is dependent on temperature, pressure and composition.  A large debate on the behavior of water, especially in the supercooled region, is currently underway [Nature Materials 2014], and due to the key role of water in biological systems a better comprehension of water structuring is desirable. 

Among the theories proposed to describe water, one model describes the extensive three-dimensional hydrogen bonded liquid water as consisting of two kind of micro-domains of different densities in dynamical equilibrium [Cho et al. 1996, Urquidi et al. 1999, Chaplin 1999, Mishima and Stanley1998, Wiggins 1995, Tanaka 2000, Wiggins 2008, Carlton 2007], namely the low-density water (LDW), with intermolecular hydrogen bonds like that of ordinary hexagonal ice, and the high-density water (HDW), with compact bonding similar to ice II. Many of the anomalies of liquid water are explained invoking a displacement of the LDW/HDW equilibrium. 
Several measurements [Chai et al. 2008, Tokushima et al. 2008, Libnau et al. 1994, Strassle et al. 2006, Errington et al. 2002] support the existence of two distinguishable structures in liquid water, especially in the supercooled region, even if they are still evident at ambient conditions where most of important biological processes occur. 

Recent experiments have shown a Debye-like slow relaxation in water, and it has been associated to structural and/or dynamical inhomogeneities on length scales quite larger than previously thought (of the order of 0.1 mm) [Jansson et al. 2010]. These observations indicate the possibility that the distorted hydrogen-bonded structures proposed by Huang et al. (2009), that dominates liquid water at ambient temperature, are connected to chainlike structures with a polymer-like dynamics. Analogous results have been obtained by spectroscopic measurements [Chai et al. 2008], showing how a structural polymorphism of water (quasi-crystalline structures) should exist also in salt solutions, where water structuring is expected, and in the solute-free zone that water forms in proximity of various hydrophilic surfaces[Chai et al. 2008]. This region has a width of 100\,$\mu$m and it is stable for days once formed.

In order to obtain further information about water structuring
we have performed Delayed Luminescence (DL) measurement on water samples. 
The term DL refers to the prolonged (from few $\mu$s up to second or minutes) photoinduced ultraweak emission of optical photons from a sample, afterwards the switching-off of the illumination source. Previously, the comparison of DL measurements performed on biological systems and solid-state systems had shown a correlation between the DL signals and the dynamic ordered structures of the samples [Scordino et al. 2000]. In particular the DL yield, i.e. the total number of photons emitted in the investigated time range, was correlated to the spatial dimension and/or to the order of the structures of the samples, while the time trends gave information on the lifetime of these structures. To explain the characteristics of the DL signals coming from several very different samples, a theoretical model was proposed, which connected the DL with the excitation and decay of non-linear coherent localized electron states (excitons or solitons) in low dimensional macromolecules [Brizhik et al. 2001] having chain-like structures and present in biological cells (i.e. alpha-helical polypeptide proteins, actin filaments, etc). It is worth noting that to explain the intrinsic luminescence of water it has been hypothesized the transmission of the excitation energy over a chain of hydrogen bonds to the luminescence emission centres, represented by defects in the structure of water [Lobyshev et al. 1999]. Moreover, to explain the large time-scale relaxation observed in sub-mm liquid samples, some kind of collective motion in hydrogen-bonded structures has been hypothesised [Jansson et al. 2010]. On the basis of these considerations the DL phenomenon seems to be a valid experimental tool to usefully investigate the structure of water. Indeed, previous works [Gulino et al. 2010, Musumeci et al. 2012] have demonstrated a significant increase in DL signal both from aqueous salt solutions near the room temperature (when, according to the literature, addition of a solute to water causes a displacement of the HDW/LDW equilibrium [Wiggins 1995, Nucci and Vanderkooi 2008, Sedlak 2006]) and from super-cooled bi-distilled water at ambient pressure. Moreover Arrhenius trend of the total number of emitted photons revealed a similar activation energy both in aqueous salt solutions and in the super-cooled water, as well as the time trends of the DL signals suggested the existence of structures unusually long-lasting in time, up to the microseconds range.

The dynamics of complex systems such as hydrogen-bonding liquids and their mixtures is nowadays one of most active areas of research. Recent studied have shown how both viscous liquid and water colloidal solutions (sol-gel TEOS samples, see below) present special dielectric effects (as e.g. the so-called \textsl{pipe-effect} [Capano et al. 2013]) due to the presence of spatially extended and long-lasting structures. The sol-gel process leads to the formation of self-assembled (nano) layers on the material surface, and this feature is used in the synthesis of new coatings with high degree of homogeneity at molecular level and outstanding physical-chemical properties. The sol-gel technique is a versatile synthetic route through which new materials 
can be obtained [Sakka 2003]. The sol-gel process is based on a two-step reaction (namely hydrolysis and condensation), starting from (semi-)metal alkoxides (e.g. usually tetraethoxysilane, $\gamma$-glycidyloxypropyltrimethoxysilane, titanium butoxide, aluminium isopropoxide) that leads to the formation of a three dimensional silica network at or near room temperature. The structure of silica phase is depending on several parameters such as type of (semi)metal precursor [Osterholz and Pohl 1992], solution pH, type and concentration of  catalyst used for hydrolysis [Didier et al. 2008, Nogami and Moriya 1980], reaction temperature, water/alkoxide ratio, heating time, nature of the alkyl chain group attached, solvent [Artaki et al. 1986] used for this process. As a consequence, the combination of these parameters determines the morphology of the resulting oxidic networks. The sol-gel based coatings are capable to protect the polymer surface by creating a physical barrier acting as insulator, thus improving the ordinary performances of the treated materials, such as flame retardancy, antimicrobial, dye fastness, anti-wrinkle finishing and biomolecule immobilization [Brancatelli et al. 2011, Alongi et al. 2013, Mahltig and Textor 2006, Mahltig and Textor 2010]. Recently, sol-gel has been also studied for innovative applications regarding hydrogen production by water photosplitting [Bhosale et al. 2012] and the development of textile materials with self-cleaning [Colleoni et al. 2012], water repellency [Colleoni et al. 2013] and sensing properties [Caldara et al. 2012, Van der Schueren et al. 2012], using microporous silica gel.

In this work we report the results of DL measurements performed on gel samples of TEOS 0.1M. The DL induced by UVA laser light was revealed in the 400-800 nm spectral range. The temporal decay was registered starting 10\,$\mu$s after the illumination pulse. We compared the responses of samples 
at different times starting from preparation (natural aging). Measurements were also performed on changing the temperature from room condition to the supercooled region. The temporal trends of the DL decays were modelled both as discrete sum of exponentials and in terms of compressed hyperbola decay laws. We applied such regression models to experimental data in order to estimate the main DL decay scale times. We also inferred the molecular binding activation energy for our water-TEOS samples by means of Arrhenius plots of the total number of emitted photons.

\section{Materials and Methods}

\subsection{Sols preparation}

\noi Tetraethoxysilane [TEOS, namely: Si(OC$_2$H$_5$)$_4$, $\geq98\%$ sol-gel precursor], hydrochloric acid (37.0\%, catalyst) were purchased from Sigma-Aldrich and used without further purification. The solvent used was distilled water.
Silica sols (0.1M) were synthesized by the sol-gel method using TEOS as silica precursor. Pure silica sol was prepared via hydrolysis: a mixture containing 2.28 ml of TEOS (0.01 mol), 0.8 ml of HCl (0.1M) and 96.92 ml of distilled water was stirred at room temperature or heated to reflux for 4 hours and 30 min. The hydrochloric acid was used to promote the hydrolysis of TEOS and the TEOS:HCl molar ratio was set to 1:\,0.008.

\begin{figure}
\begin{center}
\epsfxsize=9cm %
\epsffile{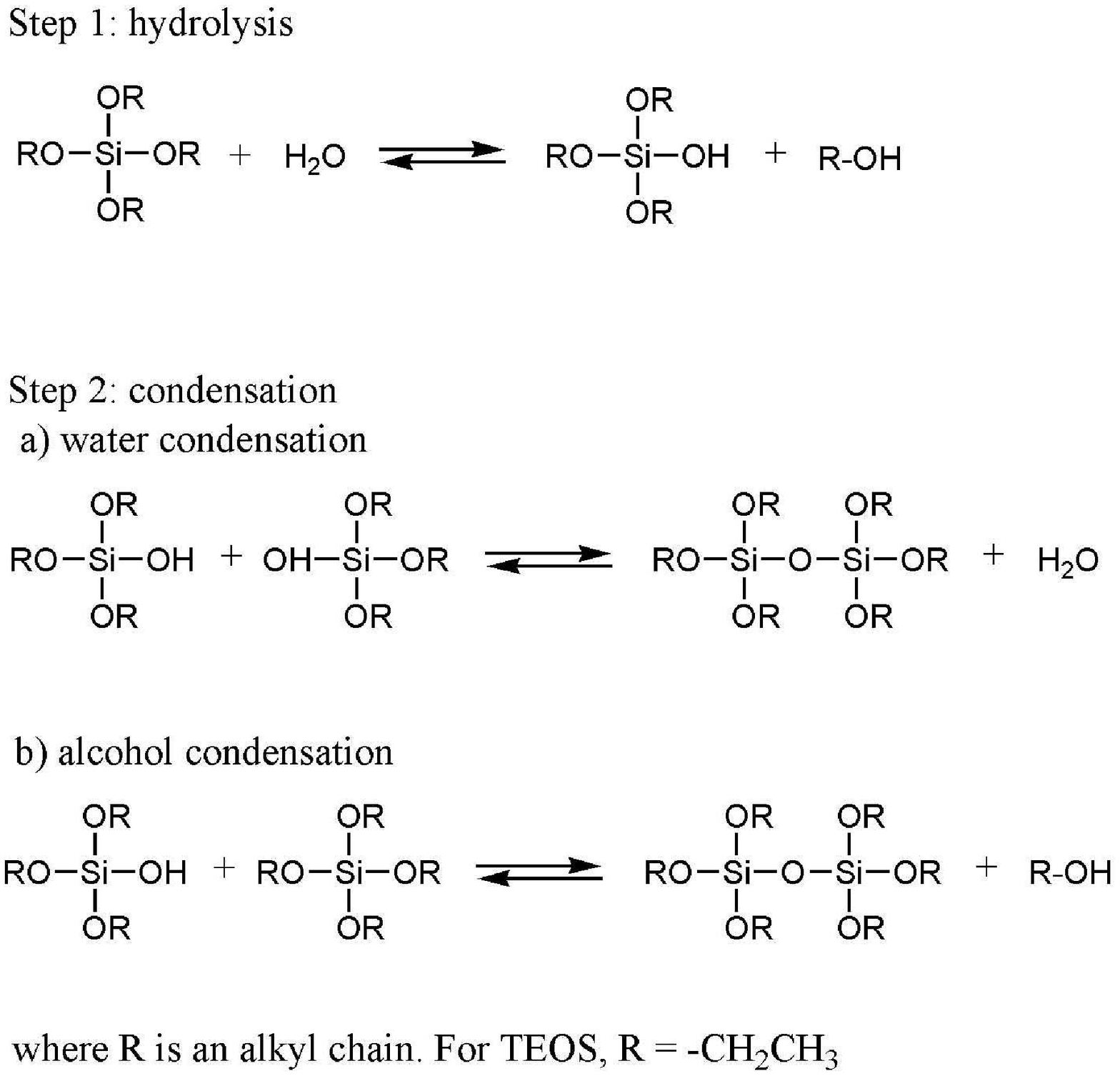} 
\vspace*{-0.3truecm}
\caption{\footnotesize Reaction scheme of the sol-gel process}
\label{Fig1}  
\lin

\end{center}
\end{figure}

\begin{figure}
\begin{center}
\epsfxsize=9cm %
\epsffile{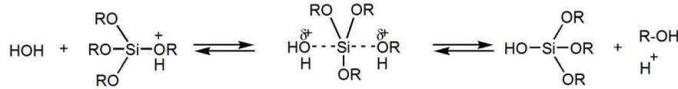} 
\caption{\footnotesize Mechanism of acid-catalysed hydrolysis {\rm S$_{\rm N}2-$Si}}
\label{Fig2}  
\end{center}
\end{figure}

Sol-gel processes are based on two steps involving hydrolysis and condensation reactions starting from (semi-)metal alkoxides. In our study, these steps occur when TEOS, water and catalyst are mixed together during the synthesis. A schematic representation of the process is reported in Fig.\,\ref{Fig1} [Alongi et al. 2011].%
Hydrolysis develops by bimolecular nucleophilic substitution involving pentacoordinate intermediates or transition states [Brinker 1988, Brinker and Seherer 1990]. The acid-catalyzed mechanisms are preceded by rapid deprotonation of alkoxide groups (-OR). The water molecule attacks from the rear and acquires a partial positive charge while the alkoxide is protonated making alcohol a better leaving group (Fig.\,\ref{Fig2}).
In this way the hydrolysis is rapid and different intermediate species containing Si-OH groups, which are called silanols, are formed. The hydrolysis products are more reactive than the un-hydrolysed precursor: consequently, further condensation reactions take place quasi-immediately after the hydrolysation.
The obtained sol becomes a gel when the solid nanoparticles dispersed in it stick together to form a network of particles spanning the liquid. This requires that the solid nanoparticles in the liquid collide with each other and stick together. This is easy for some nanoparticles since they contain reactive surface groups that make them to stick together after colliding by bonding or by electrostatic forces. However, the use of a catalyst can be required to improve the reaction rate.
Acid hydrolysis and condensation result in linear or weakly branched chains and microporous structures in silica sols, the resulting gelation times being generally long. On the opposite, uniform particles are easily formed in base catalysis, leading to a broader distribution of larger pores, which is less favourable for thermal insulation materials [Brinker 1988]. Regarding TEOS, the condensation may occur between either a silanol and a ethoxy group or two silanols to form a siloxane bond Si-O-Si by elimination of water or alcohol. When sufficient interconnected Si-O-Si bonds are realized in a region, they respond cooperatively as colloidal particles or as a sol [Buckley and Greenblatt 1994]. During the aging, the silica species react to form three-dimensional network and particle of bigger size: consequently, the viscosity of the sol increases. Nevertheless, the intensity of the viscosity depends strongly on several factors as the concentration of precursor, the presence of solvent (alcohol), the temperature of synthesis; moreover the viscosity is time dependent [Brinker 1988, Hench and West 1990]. After the sol-gel transition, the solvent phase is slowly removed from the silica network due to the additional cross-linking reactions of unreacted -OH and -OR groups.
Taking into account what above specified, 9 flasks, each about 200 ml in volume, of the same 0.1M sol were prepared and sealed at the same time and used later for DL measurements. More precisely, three flasks a time were used for each aging step. During aging before DL measurements, the flasks were stored in the dark, under controlled temperature conditions.

\subsection{Delayed Luminescence measurements}

\noi The DL measurements were performed  in the time range from 10\,$\mu$s up to 1 s by the ARETUSA set-up [Tudisco et al. 2003]. The sample was illuminated via a bifurcated optical fiber bundle (lot Oriel LLB-321) by a high-intensity pulsed nitrogen laser (Laserphotonic LN203C, $\lambda=337$\,nm, pulse width 5 ns, energy per pulse $100\pm5\,\mu$J). After the switching-off of the laser pulse, the DL signals were collected by the same fiber and revealed by a photomultiplier in single photon regime (Hamamatsu R-7206-1/Q). The photomultiplier was inhibited during the sample illumination by using an especially designed electronic shutter.  A Personal Computer implemented with a multi-channel scaler (Ortec MCS PCI) plug-in card, with a minimum dwell-time of 200 ns, was devoted to data acquisition. 
As the light absorption coefficient of water at 337 nm is very low ($<10^{-4}$\,cm$^{-1}$), a large volume of solutions (177 ml) was illuminated. The sample (TEOS solution prepared as above specified or bidistilled water from Carlo Erba) was placed inside a cylindrical stainless steel holder kept in a thermal bath (Haake C25) which allowed to reach temperatures below 0$^{\rm o}$C. The temperature was measured by a diving Pt100 temperature sensor (TP472 I.0, accuracy $\pm0.25^{\rm o}$C) connected with a thermometer (RTD HD2107.1, Delta OHM). To improve the signal to noise ratio, the head of the bifurcated optical fiber used for DL measurements was tilted of 45$^{\rm o}$C with respect to the water-air interface, so avoiding effect of internal reflection. Moreover, the fiber was immersed into the solution at the fixed distance of 3 mm from solution-air interface. 

Measurements at different temperatures were performed on cooling the sample starting from the room condition. Temperature of the coolant was lowered with a rate of about 0.5$^{\rm o}$C/min. The DL measurements were performed 20 min after the set temperature of coolant was reached, to assure thermal equilibrium inside the sample (117 ml in volume), as checked by Pt100. By using this procedure the minimum achieved temperature before water freezing was -7$^{\rm o}$C. 
Due to the low signal, no reliable spectral analysis of DL signal was performed, so the DL data refer to photons revealed in the wavelength range from 400 nm up to 800 nm corresponding to the sensitivity range of the used phototube.
Raw data were accumulated in the 50000 acquisition channels of the multi-channel scaler with a dwell time of 2\,$\mu$s, since at the end of this interval the intensity of the emitted luminescence reached values comparable with the background data. To increase the signal to noise ratio, the counts of 6000 repetitions (9000 in the case of bi-distilled water samples) of the same run were added, with a laser repetition rate of 1 Hz. To reduce random noise, a standard smoothing procedure [Scordino et al. 1996] was used, by sampling the experimental points (channel values) in such a way that final data resulted equally spaced in a logarithmic time axis. DL intensity $I(t)$ was expressed as the number of photons recorded within a certain time interval divided to the time interval and the number of repetitions. Final data are reported as average of three independent runs performed on different samples of the same type.

\subsection{Modelling Delayed Luminescence Decays}

\noi The analysis of time resolved luminescence data are usually used to get information on the structure and dynamics of the emitting system. The time relaxation of complex systems are usually described in terms of convolution of a continuous distribution of decay times as follows [Frauenfelder et al. 1999, Berberan-Santos and Valeur 2007]:
\be
I(t) = \int_0^\infty A(\tau)\erm^{-t/\tau}\rd\tau
\label{eq1} 
\ee
where the distribution function $A(\tau)$ is non-negative for all $\tau>0$ and can be regarded as a probability density function in the normalized condition $I(0)=1$. 

We used two different fitting procedures of the experimental data. The first describes the luminescence decay curve by a sum of a large number of exponentials decays and the pre-exponential terms represent discrete values of the $A(\tau)$ function. The second accords experimental data to the compressed hyperbola curve, largely used from the original work by Becquerel [Becquerel 1867] to describe the time evolution of luminescence after a short illumination [Bereran-Santos et al. 2005], and uses the inverse Laplace transform to get information on the rate constants distribution.

\

\

\

1. {\em Discrete sum of exponential decays}\\

\noi A well-known regression technique is based on a multi-exponential fit [James and Ware 1986, Branco et al. 2005]:
\be
I(t) = I_0\sum_{i=1}^Na_i\erm^{-t/\tau_i}   
\label{eq2}                                                     
\ee
where $I_0=I(0)$ and $\sum_{i=1}^Na_i=1$.
The estimate of the amplitudes $a_i$ and the lifetimes $\tau_i$ appearing in the above equation depends on the number $N$ of discrete terms used in the sum. We used two different procedures to fit experimental data, both by using Microsoft Excel, 
and in particular the \textit{Solver} function. 
In the first procedure we inferred a fitting curve with 10 exponentials and performed the non-linear fit by the iteration protocol employed by the Solver function which is based on the robust and reliable \textit{Generalized Reduced Gradient} (GRG) method [Ladson et al. 1978, Smith and Lasdon 1992]. To guide the choice of some appropriate initial conditions for amplitudes and lifetimes, we used an ordinary iterative reconvolution method, based on the Marquardt algorithm and implemented in Matlab [Nielsen 2010], involving up to four exponentials.
In the second approach the \textit{Exponential Series Method} (ESM) [Branco et al. 2005] was implemented in Excel by using again the Solver tool. The lifetimes were initially fixed with uniform distribution in logarithmic scale in an appropriate range, and the amplitudes $a_i$ were determined by linear regression, with the constraint that they had to be not negative. 
We fixed $N=15$ values of lifetimes ranging from $10^{-8}$\,s to $10^{-2}$\,s.
For the optimization of the parameters we applied some different criteria: minimization of 1) least squares, 2) $\chi^2$, 3) modified $\chi^2$, 4) mean value between maximum and minimum absolute deviations. The iterative algorithm implemented in Solver allowed us to choice one of the above criteria: namely, the one which entailed the minimum difference between the experimental data and the multi-exponential fit. Actually, we derived the fitting functions as weighted sums of 15 exponentials by imposing the minimization of the modified $\chi^2$.

\

2. {\em Compressed hyperbola decay curve}\\

\noi As for a great variety of photophysical phenomena [Berberan-Santos et al. 2005, Whitehead et al. 2009], DL temporal trend can be 
modelled by a hyperbolic function as [Scordino et al. 1996, Musumeci et al. 2005a]:
\be
I(t) = \frac{I_0}{\left(1 + \dis\frac{t}{t_0}\right)^m}
\label{eq3}
\ee                                                             
where $I_0$, $t_0$ and $m$ can be determined by using a non-linear least square fitting procedure. Indeed Eq.\,(\ref{eq3}) becomes 
linear if at fixed $t_0$ value we consider $\log(I)$ as a function of $\log(1+t/t_0)$. Starting from an initial trial value of $t_0$, 
an iterative procedure that minimize $\chi^2$ can be used [Scordino et al. 1996]. 
In terms of rate constants Eq.(1) can be written as:
\be
I(t) = I_0\int_0^\infty p(\gamma)\erm^{-\gamma t}{\rm d}\gamma
\ee
with $\int_0^\infty p(\gamma){\rm d}\gamma = 1$, so that $I(t)$ is the Laplace transform of the distribution function $p(\gamma)$ 
which can be in its turn determined by the inverse Laplace transform of $I(t)$. Such an inversion can be performed analytically 
if the observed data can be fitted by the function given by Eq.(\ref{eq3}). Indeed the Becquerel decay law possesses a simple 
inverse Laplace transform: namely a Gamma distribution [Musumeci et al. 2005a] characterized by average rate constant 
$\langle\gamma\rangle = m/t_0$ and variance  $\sigma^2=m/t_0^2$. 
Actually, it is not always possible to accord the experimental data to a single hyperbolic trend, especially if a wide 
temporal interval of the decay is considered [Musumeci et al. 2005b] and a sum of a few Becquerel functions, appropriately 
weighted, could be a powerful fitting function for complex decays [Berberan-Santos et al. 2005]. In our case, the data were 
better fitted by a bimodal function [Librizzi et al. 2002] according to the equation
\be
I(t) = \dis\frac{I_{0_1}}{\left(1 + \dis\frac{t}{t_{0_1}}\right)^{m_1}} 
+ \dis\frac{I_{0_2}}{\left(1 + \dis\frac{t}{t_{0_2}}\right)^{m_2}}
\label{eq5}
\ee                                            
such that the sum of two Gamma distributions represented the rates distribution function $p(\gamma)$ characterized by average rate constant [Berberan-Santos et al. 2005]:
\be
\langle\gamma\rangle =\frac{1}{\dis I_{0_1} + I_{0_2}}\left(I_{0_1}\frac{m_1}{t_{0_1}} + I_{0_2}\frac{m_2}{t_{0_2}}\right)
\label{eq6}
\ee

\section{Results}

\noi The aqueous solutions of TEOS, at concentration 0.1 M, were studied following the natural ageing under controlled conditions. Actually, 
the DL measurements were performed 1 week, 6 weeks and 12 weeks after solution preparation, and changing the sample temperature, i.e., at $20\gr$C, $8\gr$C, $0\gr$C, $-2\gr$C, $-4\gr$C, respectively. At each aging time the kinematic viscosity was determined by a glass capillary viscometer, in order to verify and test the formation of the silica network (see Sect.\,2). An increase of about 30\% in viscosity characterized the twelve weeks (12\textit{w)} sample with respect the one week (1\textit{w}) sample, as the viscosity value respect to the bidistilled water changed from $1.092\pm 0.008$ to $1.414\pm 0.001$.
On increasing the viscosity the samples froze at higher temperatures, since the formation of structures in the liquid favoured the silica networking, so that we were able to measure the DL of 1\textit{w}-samples down to $-4\gr$\,C, whilst the 12\textit{w}-sample were tested only down to $-2\gr$\,C.

Figure 3 shows the DL temporal trends of the 1\textit{w}-sample (Fig.\,3a) and of the 12\textit{w}-sample (Fig.3\,b) at the highest and lowest temperatures, respectively, after background subtraction. The markers denote the average values of measurements corresponding to three different samples of the same type, while vertical bars refer to one standard deviation. It is evident that the DL intensity increases on cooling, as previously observed both in aqueous salt solutions and in bidistilled water [Gulino et al. 2010, Musumeci et al. 2012].

\begin{figure}
\begin{center}
\epsfxsize=9cm %
\epsffile{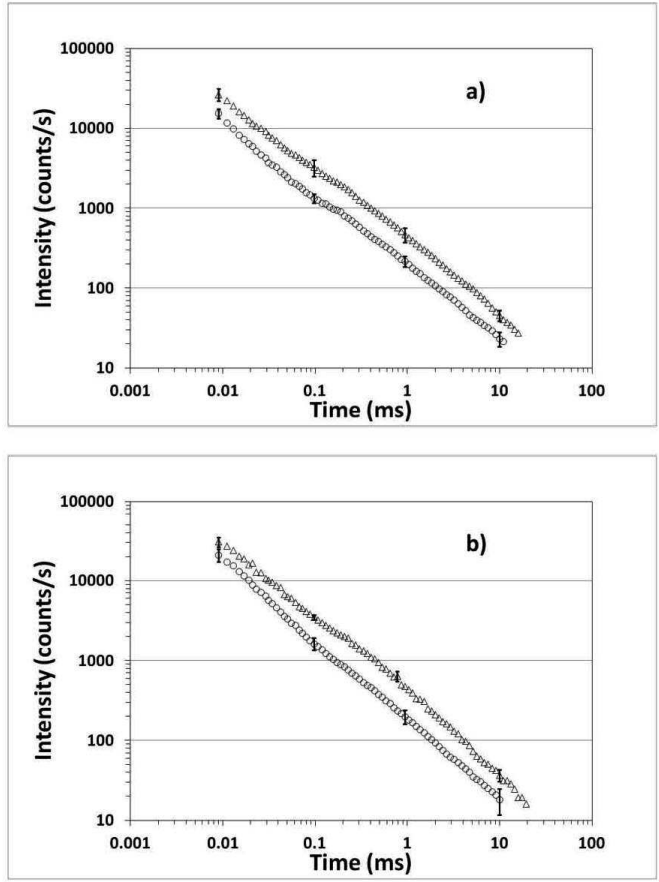} 
\end{center}
\caption{\footnotesize Time trend of the DL signals from 1\textit{w}-sample (a) and 12\textit{w}-sample (b) at the highest and lowest temperatures; $(\circ$) corresponds to $T_{\rm max} = 20\gr$; ($\triangle$) corresponts to $T_{\rm min} = -4\gr$\,C in (a) and to $T_{\rm min} = -2\gr$\,C in (b). Markers denote average values of three different samples; vertical bars denote the standard deviation at pre-fixed times of the decay.}
\label{Fig3}  
\end{figure}

\

\noi The evaluation of the total number $N_{\rm counts}$ of DL counts, in the time windows of the decay, follows from the experimental intensity data $I_{\rm exp}(t)$ as:
\be
N_{\rm counts} = \int_{t_i}^{t_f}I_{\rm exp}(t)\rd t
\ee                                                                
where $t_i$ and $t_f$ are the initial and final experimental times of the observed decay, respectively. The experimental data were integrated numerically, and in Fig.\,4 we report the logarithm of $N_{\rm counts}$ as a function of the inverse of the temperature for the 1\textit{w}- and 12\textit{w}-samples. The experimental points follow a linear trend, thus denoting an Arrhenius behavior characterized by a well-defined activation energy $\Delta E$. The linear fit gives $\Delta E = (22\pm 2)$\,kJ/mol (with $R^2 = 0.978$) in the case of 1\textit{w}-sample, and $\Delta E = (25\pm 2)$\,kJ/mol (with $R^2 = 0.990$) in the case of 12\textit{w}-sample.
 
\begin{figure}
\begin{center}
\epsfxsize=9cm %
\epsffile{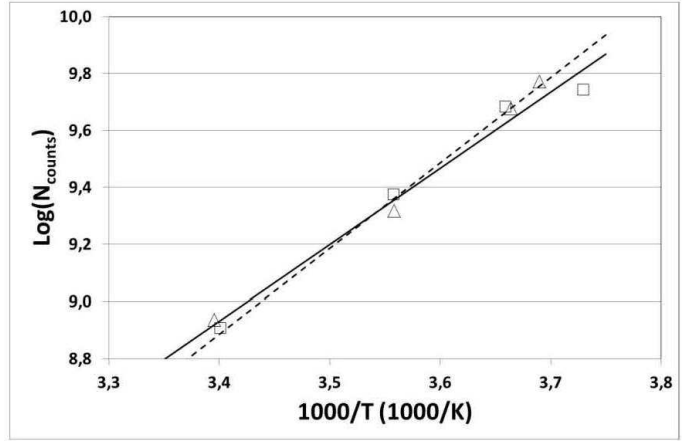} 
\end{center}
\caption{\footnotesize Arrhenius plot of the total number $N_{\rm counts}$ of DL emitted photons: ($\Box$) 1\textit{w}-sample, ($\triangle$) 12\textit{w}-sample (the error bars of the average values are smaller than the marker sizes). Straight lines corresponding to the linear fits are showed as well: 1\textit{w}-sample (solid line), 12\textit{w}-sample (dashed line).}
\label{Fig4}  
\end{figure}

\noi The temporal analysis of the experimental data was made, as specified in Sect.\,2, applying two different fitting procedures to the time relaxation of complex systems as convolution of continuum distribution of exponential decays.
By using the regression technique based on the abovementioned multi-exponential fit for any measurement campaign and any experimental data set we found (see Fig.\,5) a very large amplitude for $\tau\simeq 5\,\mu$s, followed by much smaller ones related to $\tau_i\simeq 2\cdot10^{-4}$\,s and $10^{-3}$\,s respectively. In Fig.\,6 we notice that the experimental data agree very well with the regression plot obtained with the 15-exponential fit.

\begin{figure}
\begin{center}
\epsfxsize=9cm %
\epsffile{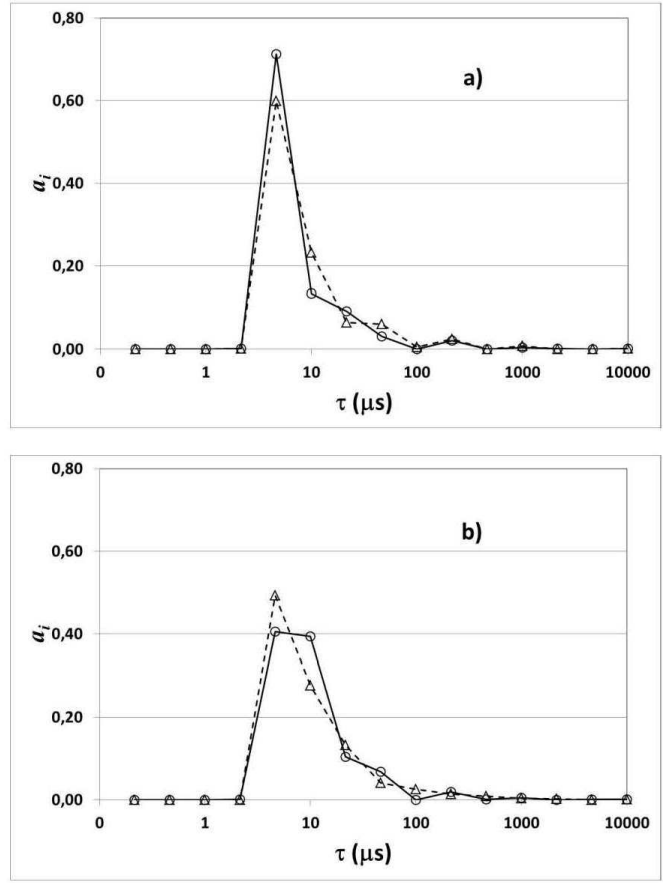} 
\end{center}
\caption{\footnotesize Pre-exponential amplitudes $a_i$ as a function of lifetimes $\tau$ in the case of 1\textit{w}-sample (a) and 12\textit{w}-sample (b) at temperatures $T = 20\gr$C ($\circ$) and $T = 0\gr$C ($\triangle$).}
\label{Fig5}  
\end{figure}

\begin{figure}
\begin{center}
\epsfxsize=9cm %
\epsffile{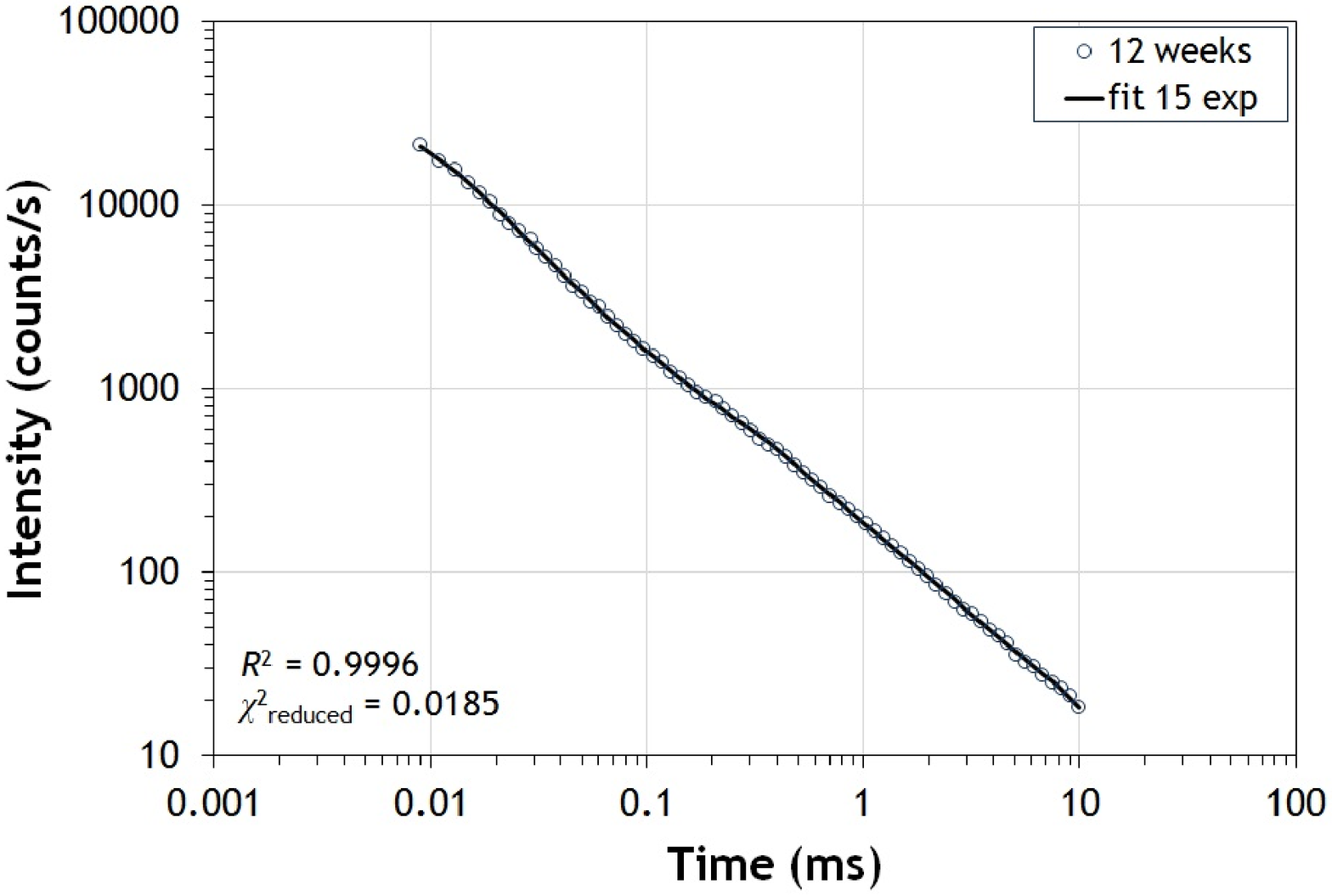} 
\end{center}
\caption{\footnotesize DL signals versus time for $12w$-sample at  $T = 20\gr$C. The fitting curve (solid line) is given by Eq.\,(\ref{eq2}) with $N=15$.}
\label{Fig6}  
\end{figure}
 
\noi Alternatively, by according the data to the compressed hyperbola decay curve Eq.\,(\ref{eq5}), the fitting parameters $I_{0i}$, $t_{0i}$ and $m_i$ were evaluated as described in Sect\,2. Interestingly, we found minimum reduced-$\chi^2$ figures for $t_{01}$ and $t_{02}$ equal to about 0.01 and 0.1\,ms, respectively. So we set $t_{01} = 0.01$\,ms and $t_{02} = 0.1$\,ms for any sample in every condition (obtaining a reduced $\chi^2\leq 1$); the fit parameters are reported in Table I.%
\begin{figure}
\lin
\begin{center}
\epsfxsize=12cm %
\epsffile{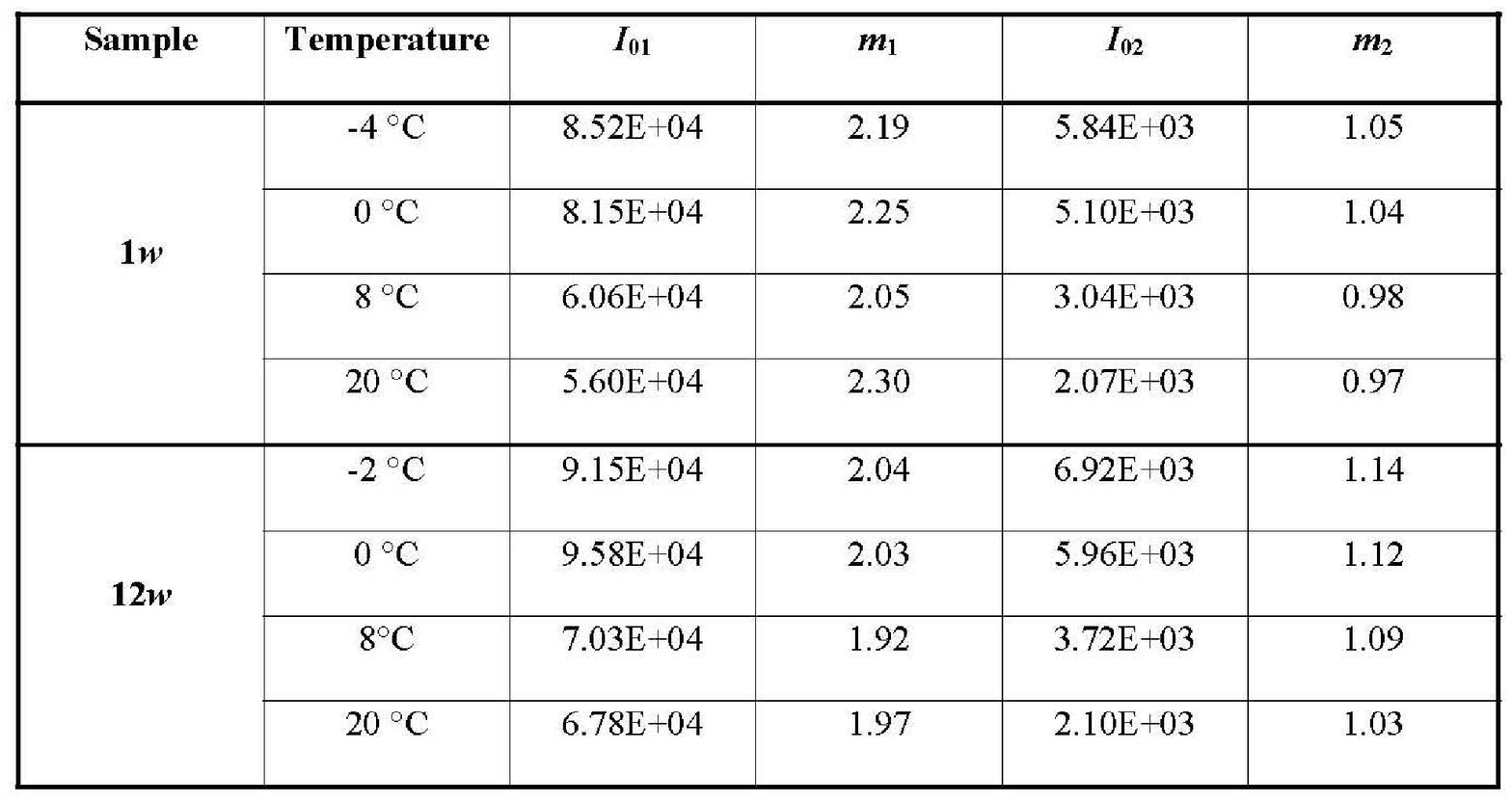} 
\end{center}
\vspace*{-0.3cm}
{\footnotesize TABLE I: Fitting parameters of the compressed hyperbola decay curve Eq.(5), with $t_{01}= 0.01$\,ms and $t_{02}=0.1$\,ms (reduced $\chi^2$ is $\leq1$).}
\label{TABLE1}  
\end{figure}
\noi The average rate constant Eq.\,(\ref{eq6}) of each decay resulted to be 
\mbox{$(2.1\pm 0.1)\,10^5$\,\,s$^{-1}$} and \mbox{$(1.89\pm 0.04)\,10^5$\,\,s$^{-1}$} for the 1\textit{w} and 12\textit{w} samples respectively. As expected, these rate constants values
well agree with the lifetimes obtained with the multi-exponential fit.
The relative errors for $m_1$, $m_2$, $I_{02}$ are generally quite small (lower than 3\%, 1\%, 7\%, respectively), while $I_{01}$ is affected by a larger error (not lower than 35\%), even if a visual inspection of the fitted curve shows a good agreement between theoretical curves and experimental points. The large error in $I_{01}$ evaluation is due to the fact that the experimental points are actually at times longer than $t_{01}$, which coincides with the first experimental temporal point: as a consequence the behavior of the decay at shorter times can be only inferred with a greater uncertainty.
To overcome this problem, according to the theoretical trend given by Eq.\,(\ref{eq5}), we tried to get further information by using the ``theoretical points'' that are better resolved: i.e., those related to longer times in 
\be
I_{2_{\rm th}}(t) = \frac{I_{0_2}}{\dis\left(1+\frac{t}{t_{0_2}}\right)^{m_2}}
\ee
Therefore we evaluated numerically the contribution to the total number of emitted photons $N_{\rm counts}$ coming from the shorter decay times emitters ($N_1$) and from the longer decay times emitters ($N_2$) as follows
\be
N_1 =\int_{t_i}^{t_f}\left[I_{\rm exp}(t) - I_{2_{\rm th}}(t)\right]\rd t
\ee
\be
N_2 = N_{\rm counts} - N_1
\ee
In Fig.\,7 $N_1$, $N_2$ and $N_{\rm counts}$ are reported as a function of the sample temperature for the two different aging conditions. 

\begin{figure}
\begin{center}
\epsfxsize=9cm %
\epsffile{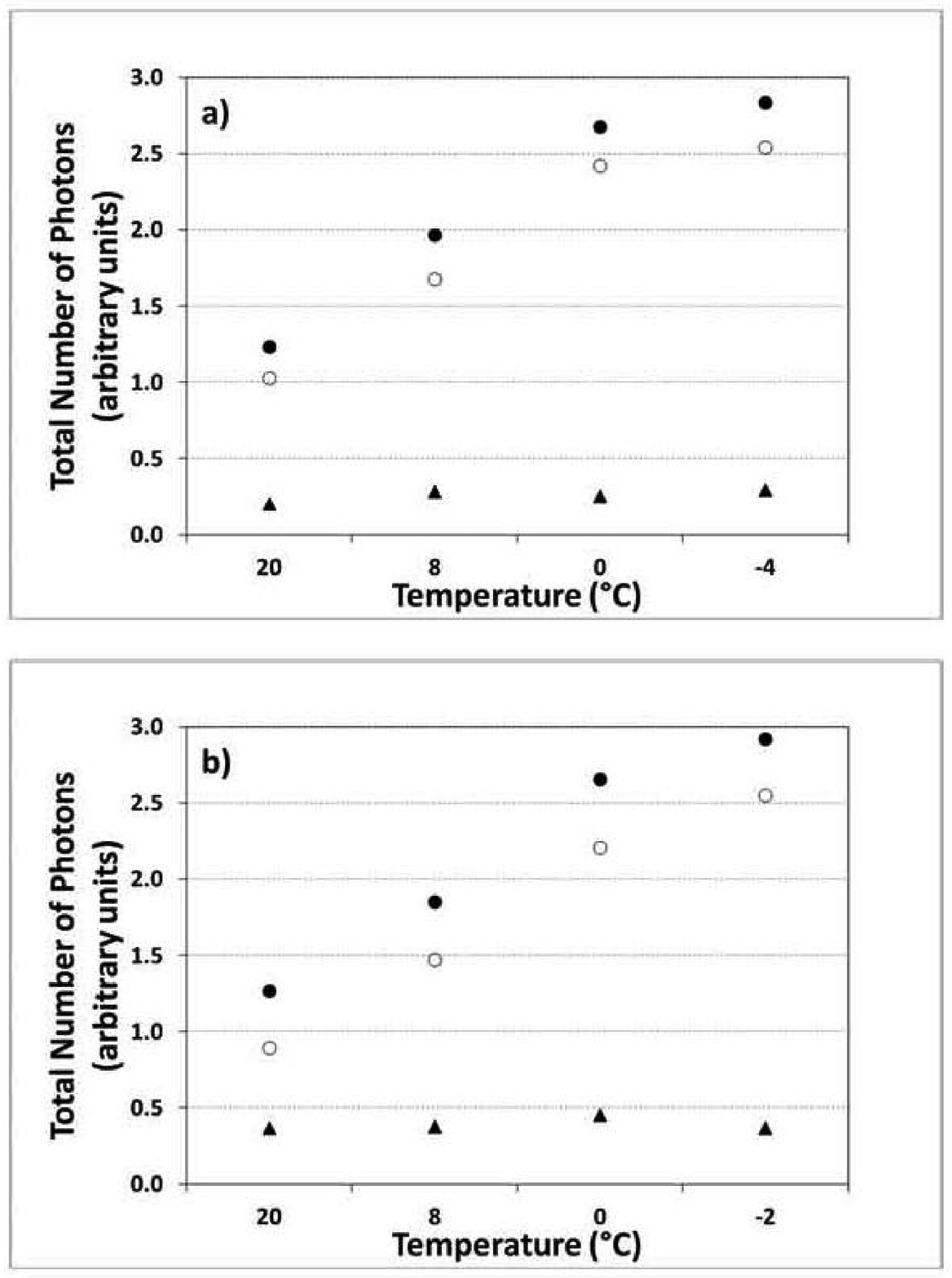} 
\end{center}
\caption{\footnotesize Total number of emitted photons in the whole decay time $N_{\rm counts}$ ($\bullet$), from shorter decay times emitters $N_1$ ($\blacktriangle$) and from longer decay times emitters $N_2$ ($\circ$)as a function of the temperature of the sample. (a) 1\textit{w}-sample; (b) 12\textit{w}-sample. Data are the average values of three different experiments (standard deviation are within marker size).}

\lin
\label{Fig7}  
\end{figure}

\noi The largest contribution to DL intensity comes from the decay of excited states with longer lifetimes (of the order of milliseconds) $N_2$ which increases on decreasing the sample temperature, while the contribution from shorter lifetimes excited states $N_1$  is quite constant with temperature, with an increase of about 45\% after aging.

The Arrhenius plots of $N_2$ give an activation energy of $\Delta E = (24\pm 3)$\,kJ/mol ($R^2 = 0.976$) and $\Delta E = (31\pm 2)$\,kJ/mol ($R^2 = 0.993$) for the 1\textit{w}- and 12\textit{w}-sample, respectively, to be compared with the above reported values. The activation energy for $N_2$ almost coincides (within the error bars) with that for $N_{\rm counts}$ for the fresh samples, whilst it is larger for the aged one, with more pronounced structures. No significant linear correlation of $\log N_1$ with inverse of the temperature was found.

\begin{figure}
\begin{center}
\epsfxsize=9cm %
\epsffile{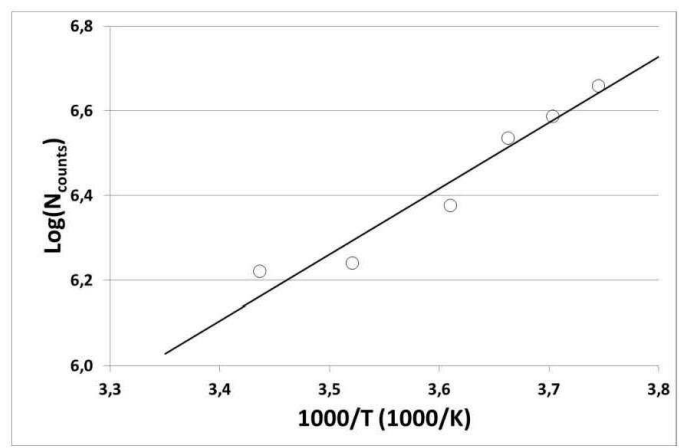} 
\end{center}
\caption{\footnotesize Arrhenius plot of the total numbers Ncounts of DL photons emitted by bi-distilled water. Markers represent average values whose standard deviations are inside the marker size. The line represent fit according to an Arrhenius trend ($R^2=0.940$).}
\label{Fig8}  
\end{figure}

For comparison, in Fig.\,\ref{Fig8} we report the DL behavior in bi-distilled water as a function of temperature for which we get an activation energy $\Delta E = (13\pm 2)$\,kJ/mol ($R^2 = 0.940$).%
Actually, the fitting parameters obtained for the DL time decays of pure water gave reasonable values also for $t_{01}$ of order of 1\,$\mu$s. Moreover, in agreement with the results obtained, no significant correlation of $N_1$ with temperature was observed, while a poor correlation ($R^2 = 0.865$) of $\log N_2$ with temperature (data not shown) gave  $\Delta E = (16\pm 3)$\,kJ/mol, equal within the errors to the value for $N_{\rm counts}$. 

\section{Discussion}

\noi As observed in previous analyses, DL is able to investigate complex structures formed in liquids such as water.  Such structures have been easily observed in salt solutions of water, where the ionic compounds evidently favour domain formation. Here we have observed a similar dynamics induced by TEOS, where the gelation process leads to the formation of a three-dimensional silica network which involves alcohol and water elimination according to the chemical reaction reported in Fig.\,1. The total number of photons collected from aqueous solutions of TEOS shows, again, an Arrhenius behaviour, but with a larger activation energy, both with respect to pure water and to salt-water solutions [Gulino et al. 2010, Musumeci et al. 2012]. In particular, the activation energy of about 13 kJ/mol for bi-distilled water increases up to about 25 kJ/mol for 12\textit{w} water-TEOS solutions. This energy value, typically required for breaking and completely separating hydrogen bonds, results to depend quite appreciably on the solution aging, which directly affects the number of photon-sensitive domains created in the gelation process. Moreover, the increasing value of the activation energy with aging for the long-lasting excited states $N_2$ (up to 31 kJ/mol for 12\textit{w} water-TEOS solutions) could be related to the formation of larger clusters of water molecules, promoted by the gelation process. Actually, a similar increasing in the binding energy per molecule and in the fragmentation energy with increasing the water cluster size has been predicted in a recent ab initio calculation studying the stability of water clusters [Liu et al. 2011]. Interestingly the same percentage increase (about 29\%) with aging occurs both in $\Delta E$ (for $N_2$) and in viscosity.

However, the most striking results come from the lifetime analysis of the luminescence decays. Indeed, we have found that, independently from  temperature and aging conditions, in the DL lifetime spectrum the maximum amplitude $A$ comes from domains endowed with a lifetime $\tau_0$ of about 5\,$\mu$s. This likely refers to a true physical property of water domains, since the same lifetime is observed both in bi-distilled water (data not shown), in salt solution [Gulino 2010] and in  silica gel solutions (see Fig. \ref{Fig5}).
 
Moreover, information about the interaction of TEOS with water can be gained from the measurements of the DL amplitude $A$ for water domains with lifetime $\tau_0$: $A$ in fact depends significantly on the aging conditions of the sample, decreases with increasing aging, and moves to a slight longer lifetime. Accordingly, the average rate constant (Eq.(6) and Table I) decreases with aging.

In addition, the lifetime analysis evidences different lifetimes $\tau_i$, 
much higher than the intrinsic value $\tau_0$,  although with subdominant DL activities ($a_i \ll A$) that could be related to the formation of different domains that contribute to the long-lasting excited states $N_2$. Contrary to the above intrinsic feature, such secondary lifetimes  are driven by the presence of the substance dissolved in water (i.e. TEOS), and depend 
on the temperature of the sample, as well as on the aging conditions, thus denoting a characteristic property of the solute rather than of water. Quite intriguingly, we see that a large aging of TEOS solution, which increases silica networking with consequent water elimination, nevertheless induces a large DL from water domains even with a lesser effective content of water molecules. We can infer that the larger presence of formed silicates strongly favours the aggregation of the remaining water molecules in structured domains, thus compensating the diminution of water content and even increasing the number of DL-sensitive aggregates.

Summarizing, the measurements reported in this work 
not only prove the existence of intrinsic DL for water, thus allowing us to estimate its lifetime, but also provide new important information about structure and dynamics of water domains, including the aggregation strength measured by the Arrhenius activation energy and the intensity and features of solute-water interaction. 

\

\begin{acknowledgments}

\noi The authors acknowledge prof.\,Giovanna Di Pasquale (Dipartimento di Ingegneria Industriale, Universit\`a di Catania, Italy) for providing viscosity measurements of the samples.

\end{acknowledgments}



\begin{thebibliography}{}

\bibitem{1} J. Alongi, M. Ciobanu and G. Malucelli , Cellulose {\bf 18} (2011) 167

\bibitem{2} J. Alongi, C. Colleoni, G. Rosace and G. Malucelli, Cellulose {\bf 20} (2013) 525

\bibitem{3} I. Artaki, T.W. Zerda and J. Jonas, J. Non-Cryst. Sol. {\bf 81} (1986) 381

\bibitem{4} E. Bequerel, \textit{La lumire: ses causes et ses effets} (1867) {\bf 1} (Firmin Didot, Paris)

\bibitem{5} M.N. Berberan-Santos, E.N. Bodunov and B. Valeur, Chem. Phys. {\bf 317} (2005) 57

\bibitem{6} M.N. Berberan-Santos and B. Valeur, J. Lumin. {\bf 126} (2007) 263

\bibitem{7} R.R. Bhosale, R.V. Shende and J.A. Puszynski, Int. J. Hydrogen Energy {\bf 37} (2012) 2924

\bibitem{8} G. Brancatelli, C. Colleoni, M.R. Massafra and G. Rosace, Polym. Degrad. Stabil. {\bf 96} (2011) 483

\bibitem{9} T.J.F. Branco, A.M. Botelho do Rogo, I.F. Machado and L.F. Vieira Ferreira, J Phys Chem {\bf B109} (2005) , 15958

\bibitem{10} C.J. Brinker, J. Non-Cryst. Sol. {\bf 100} (1988) 31

\bibitem{11} C.J. Brinker and G.W. Seherer, \textit{Sol-Gel Science: The Physics and Chemistry of Sol-Gel Processing} (Academic Press, San Diego; USA, 1990)

\bibitem{12} L. Brizhik, A. Scordino, A. Triglia and F. Musumeci, Phys. Rev. {\bf E64} (2001) 031902

\bibitem{13} A.M. Buckley and M. Greenblatt, J. Chem. Educ. {\bf 71} (1994) 599

\bibitem{14} M. Caldara, C. Colleoni, E. Guido, V. Re and G. Rosace, Sensor Actuat. Chem. {\bf B171} (2012) 1013

\bibitem{15} V. Capano, S. Esposito and G. Salesi, Eur. Phys. J. Appl. Phys. {\bf 62} (2013) 31103 

\bibitem{16} T.S. Carlton, J. Phys. Chem.  {\bf B111} (2007) 13398

\bibitem{17} B.H. Chai, J.M. Zheng, Q. Zhao and G.H. Pollack, J. Phys. Chem. {\bf A112} (2008) 2242

\bibitem{18} M.F. Chaplin, Biophys. Chem. {\bf 83} (1999) 211

\bibitem{19} C.H. Cho, S. Singh and G.W. Robinson, Phys. Rev. Lett. {\bf 76} (1996) 1651

\bibitem{20} C. Colleoni, M.R. Massafra and G. Rosace, Surf Coat Tech {\bf 207} (2012) 79

\bibitem{21} C. Colleoni, E. Guido, V. Migani and G. Rosace, J. Industrial Textile, in press (2013) DOI: 10.1177/1528083713516664

\bibitem{22} B. Didier, R. Mercier, N.D. Alberola and C. Bas, J. Polym. Sci. Pt. Polym. Phys. {\bf B46} (2008) 1891

\bibitem{23} J.R. Errington, P.G. Debenedetti and S. Torquato, Phys. Rev. Lett. {\bf 89} (2002) 215503

\bibitem{24} H. Frauenfelder, P.G. Wolynes and R.H. Austin, Rev. Mod. Phys. {\bf 71} (1999) S419

\bibitem{25} M. Gulino, R. Grasso, L. Lanzan, A. Scordino, A. Triglia, S. Tudisco and F. Musumeci, Chem. Phys. Lett. {\bf 497} (2010) 99 

\bibitem{26} L.L. Hench and J.K. West, Chem. Rev. {\bf 90} (1990) 33

\bibitem{27} C. Huang, K.T. Wikfeldt, T. Tokushima, D. Nordlund, Y. Harada, U. Bergmann, M. Niebuhr, T.M. Weiss, Y. Horikawa, M. Leetmaa, M.P. Ljungberg, O. Takahashi, A. Lenz, L. Ojam\"{a}e, A.P. Lyubartsev, S.Shin, L.G.M. Pettersson and A. Nilsson, PNAS {\bf 106} (2009) 15214

\bibitem{28} D.R. James and W.R. Ware, \textit{Recovery of underlying distributions of lifetimes from fluorescence decay data} (Photochemistry Unit, Department of chemistry, The University of Western Ontario; Canada, 1986)

\bibitem{29} H. Jansson, R. Bergman and J. Swenson, Phys. Rev. Lett. {\bf 104} (2010) 017802

\bibitem{30} L.S. Lasdon, A.D. Waren, A. Jain and M. Ratner, ACM Trans. Mathematical Software {\bf 4} (1978) 34

\bibitem{31} F.O. Libnau, J. Toft, A.A. Christy and O.M. Kvalheim, J. Am. Chem. Soc. {\bf 116} (1994) 8311

\bibitem{32} F. Librizzi, C. Viappiani, S. Abbruzzetti and R. Cordone, J. Chem. Phys. {\bf 116} (2002) 1193

\bibitem{34} X. Liu, W.- C. Lu, C.Z. Wang, and K.M. Ho, Chem. Phys. Lett. {\bf 508} (2011) 270

\bibitem{35} V.I. Lobyshev, R.E. Shikhlinskaya and B.D. Ryzhikov, J. Mol. Liq. {\bf 82} (1999)73

\bibitem{36} B. Mahltig and T.Textor, J. Sol-Gel Sci. Techn. {\bf 39} (2006) 111

\bibitem{37} B. Mahltig and T. Textor, Fiber Polym. {\bf 11} (2010) 1152

\bibitem{38} O. Mishima and H.Stanley, Nature {\bf 396} (1998) 329

\bibitem{39} F. Musumeci, L.A. Applegate, G. Privitera, A. Scordino, S. Tudisco and H.J. Niggli, J. Photochem. Photobiol. Biol. {\bf B79} (2005) 93

\bibitem{40} F. Musumeci, G. Privitera, A. Scordino, S. Tudisco , C. Lo Presti, L.A. Applegate and H.J. Niggli, Appl. Phys. Lett. {\bf 86} (2005) 153902 

\bibitem{41} F. Musumeci, R. Grasso, L. Lanzan, A. Scordino, A. Triglia, S. Tudisco and M. Gulino, J. Biol. Phys. {\bf 38} (2012) 181

\bibitem{42} H.B. Nielsen, \textit{A Matlab toolbox for optimization and data fitting} (DTU Informatics, IMM Technical University of Denmark; 2010)

\bibitem{43} M. Nogami and Y.J. Moriya, J. Non-Cryst. Sol. {\bf 37} (1980) 191

\bibitem{44} N.V. Nucci and J.M. Vanderkooi, J. Mol. Liq. {\bf 143} (2008) 160

\bibitem{45} F.D. Osterholz and E.R. Pohl, \textit{In Silanes and Other Coupling Agents} pp. 119-141 (K. L. Mittal Ed., VSP; Utrecht 1992) 

\bibitem{46} S. Sakka,\textit{ Sol-Gel Science and Technology. Topics and Fundamental Research and Applications} (Kluwer Academic Publishers; Norwell 2003)

\bibitem{47} A. Scordino, A. Triglia, F. Musumeci, F. Grasso and Z. Rajfur, J. Photochem. Photobiol. Biol. {\bf B32} (1996) 11

\bibitem{48} A. Scordino, A. Triglia and F. Musumeci, J. Photochem. Photobiol. Biol. {\bf B56} (2000) 181

\bibitem{49} M. Sedlk, J. Phys. Chem. {\bf B110} (2006) 13976

\bibitem{50} S. Smith and L. Lasdon, ORSA J. Comput. {\bf 4} (1992) 2

\bibitem{51} H.E. Stanley, P. Kumar, G. Franzese, L. Xu, Z. Yan, M.G. Mazza, S.V. Buldyrev, S.-H. Chen and F. Mallamace, Eur. Phys. J. Special. Topics {\bf 161} (2008) 1

\bibitem{52} Th. Strssle, A.M. Saitta, Y. Le Godec, G. Hamel, S. Klotz, J.S. Loveday and R.J. Nelmes, Phys. Rev. Lett. {\bf 96} (2006) 067801

\bibitem{53} H. Tanaka, J. Chem. Phys. {\bf 112} (2000) 799

\bibitem{54} T. Tokushima, Y. Harada, O. Takahashi, Y. Senba, H. Ohashi, L.G.M. Pettersson, A. Nilsson and S. Shin, Chem. Phys. Lett. {\bf 460} (2008) 387

\bibitem{55} S. Tudisco, F. Musumeci, A. Scordino and G. Privitera, Rev. Sci. Instrum. {\bf 74} (2003) 4485

\bibitem{56} J. Urquidi, C.H. Cho, S. Singh and G.W. Robinson, J. Mol. Struct. {\bf 363} (1999) 485

\bibitem{57} L. Van der Schueren, K. De Clerck, G.Brancatelli, G.Rosace, E. Van Damme and W. De Vos, Sensor. Actuat. Chem. {\bf B162} (2012) 27

\bibitem{58} L. Whitehead , R. Whitehead , B. Valeur and M.N. Berberan-Santos, Am. J. Phys. {\bf 77} (2009) 173

\bibitem{59} P.M. Wiggins, Microbiol. Rev. {\bf 54} (1990) 432

\bibitem{60} P.M. Wiggins, Prog. Polym. Sci. {\bf 20} (1995) 1121

\bibitem{61} P.M. Wiggins, PLoS ONE {\bf 3} (2008) e1406

\bibitem{nature} Editorial, Debated waters, Nature Materials {\bf 13} (2014) 663



\end{thebibliography}
\end{document}